\documentclass{PoS}

\title{The Spectrum of tmLQCD with Quark and Link Smearing}

\ShortTitle{The Spectrum of tmLQCD with Quark and Link Smearing}

\author{\speaker{Abdou M. Abdel-Rehim}\\
        Department of Physics, University of Regina,
        Regina, SK, S4S 0A2, Canada\\
        and Department of Physics, Baylor University, Waco, TX, USA 76798-7316\\
        E-mail: \email{Abdou\_Abdel-Rehim@baylor.edu}}
\author{Randy Lewis, Robert G. Petry\\
        Department of Physics, University of Regina,
        Regina, SK, S4S 0A2, Canada}
\author{R. M. Woloshyn\\
        TRIUMF, 4004 Wesbrook Mall, Vancouver, BC, V6T 2A3, Canada}
        
\abstract{The effect of using smeared sink operators on the hadron spectrum
is studied for quenched twisted mass lattice QCD with up, down, and
strange quarks. Gaussian smearing is used for quark fields, and stout
link smearing for gauge fields. Smeared correlators are found to be
dominated by the ground state with a small contribution from excited
states, leading to an improved determination of some ground state masses.}

\FullConference{XXIVth International Symposium on Lattice Field Theory\\
                July 23-28, 2006\\
                Tucson, Arizona, USA}

\begin{document}

\section{Introduction}
\subsection{Motivation}
Twisted mass lattice QCD (tmLQCD) offers an efficient mechanism for
eliminating unphysical zero modes\cite{Frezzotti:2000nk}
%[Frezzotti, Grassi, Sint and Weisz, JHEP 08, 058 (2001)]
and removing $O(a)$ errors from Wilson simulations\cite{Frezzotti:2003ni}.
%[Frezzotti and Rossi, JHEP 08, 007 (2004)].
An advantage of this formulation is that it allows simulations to be done
at relatively light quark masses. Good results can be obtained for 
pions\cite{Abdel-Rehim:2005gz,Jansen:2005} but the correlation functions
for other hadrons tend to have significant statistical fluctuations when
the quarks are light. In particular our
previous exploration of the octet and decuplet baryons indicated that
simple local operators are inadequate at small quark
masses\cite{Abdel-Rehim:2005qv}.
%[Abdel-Rehim, Lewis and Woloshyn, PoS LAT2005:032,2006].
Here, we study whether the application of sink smearing can be useful in 
improving baryon signals and reducing statistical errors.
\subsection{Action}
We use the Wilson gauge action with a tmLQCD fermion action\cite{Pena:2004gb}
%[Pena, Sint and Vladikas, JHEP 09, 069 (2004)]
\begin{equation}
S_f[\psi,\bar\psi,U] = a^4\sum_x\bar\psi(x)\left(M+i\mu\gamma_5
      +\gamma\cdot\nabla^\pm
      -\frac{a}{2}\sum_\nu\nabla^*_\nu\nabla_\nu\right)\psi(x),
\end{equation}
where
\begin{eqnarray}
\psi^T &=& (u,d,c,s), \\
M &=& diag(M_l,M_l,M_c,M_s), \\
\mu &=& diag(\mu_l,-\mu_l,\mu_c,-\mu_s).
\end{eqnarray}
Forward ($\nabla$), backward ($\nabla^*$) and symmetric ($\nabla^\pm$)
derivatives are standard, and
the $c$ field is irrelevant for our quenched simulations.
\subsection{Simulation details}
Simulation parameters are given in Table \ref{Simulation-Parameters}.
These parameters correspond to maximal twist using the parity
definition\cite{Abdel-Rehim:2005gz,Farchioni:2004ma,Farchioni:2004fs}.
In the following, multi-state fits are used to extract ground state masses.
\begin{table}[b]
\begin{center}
\begin{tabular}{cccccl}
\hline
$\beta$ & \#sites & \#configs & $aM$ & $a\mu$ & physical mass \\
\hline
5.85 & $20^3\times40$ & 594 & -0.8965 & 0.0376 & $\sim m_s$ \\
     &                &     & -0.9071 & 0.0188 & $\sim m_s/2$ \\
     &                &     & -0.9110 & 0.01252 & $\sim m_s/3$ \\
     &                &     & -0.9150 & 0.00627 & $\sim m_s/6$ \\
6.0 & $20^3\times48$ & 600 & -0.8110 & 0.030 & $\sim m_s$ \\
    &                &     & -0.8170 & 0.015 & $\sim m_s/2$ \\
    &                &     & -0.8195 & 0.010 & $\sim m_s/3$ \\
    &                &     & -0.8210 & 0.005 & $\sim m_s/6$ \\
6.2 & $28^3\times56$ & 200 & -0.7337 & 0.021649& $\sim m_s$ \\
    &                &     & -0.7367 & 0.010825& $\sim m_s/2$ \\
    &                &     & -0.7378 & 0.007216& $\sim m_s/3$ \\
    &                &     & -0.7389 & 0.003608& $\sim m_s/6$ \\
\hline
\end{tabular}
\end{center}
\caption{Simulation parameters.}
\label{Simulation-Parameters}
\end{table}

\section{Smearing Method} 
\begin{figure}
\begin{center}
\scalebox{0.6}{\includegraphics*[10mm,26mm][25cm,185mm]{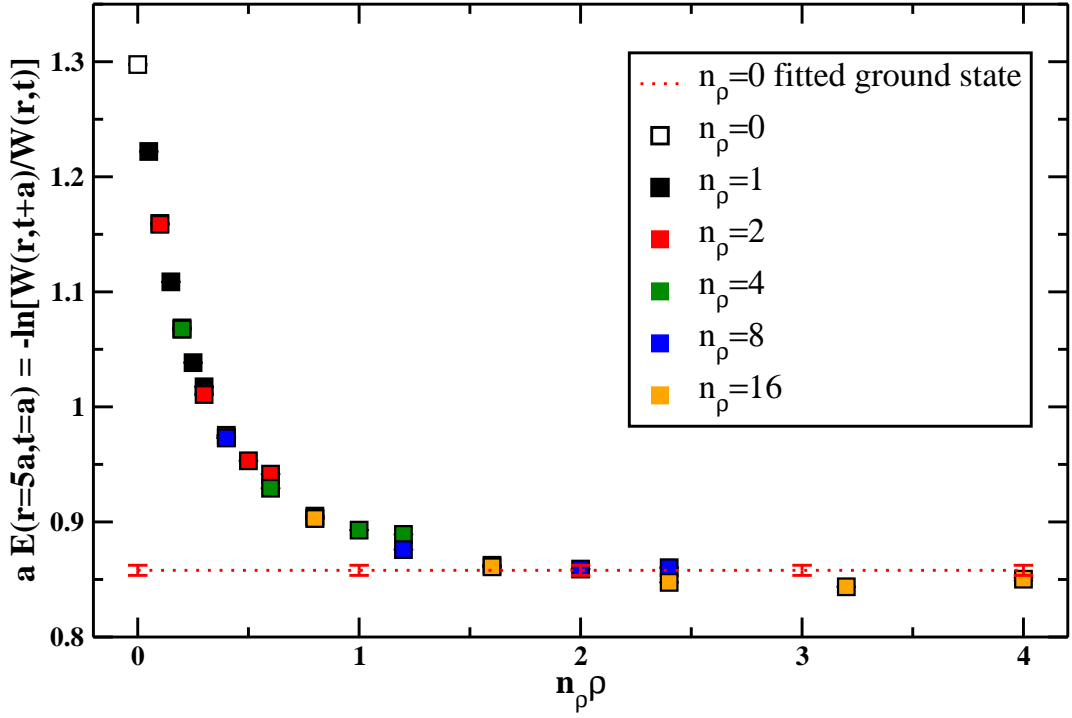}}
\end{center}
\caption{Effective mass for $\beta=6.0$ at temporal separation t=a
corresponding to the static quark-antiquark potential
at spatial separation r=5a,
as a function of the stout link smearing parameters $\rho$
and $n_\rho$.
The fitted ground state shown with no smearing used a two-exponential fit.}
\label{qqbar-potential}
\end{figure}

\begin{figure}
\begin{center}
\scalebox{0.6}{\includegraphics*[10mm,19mm][25cm,185mm]{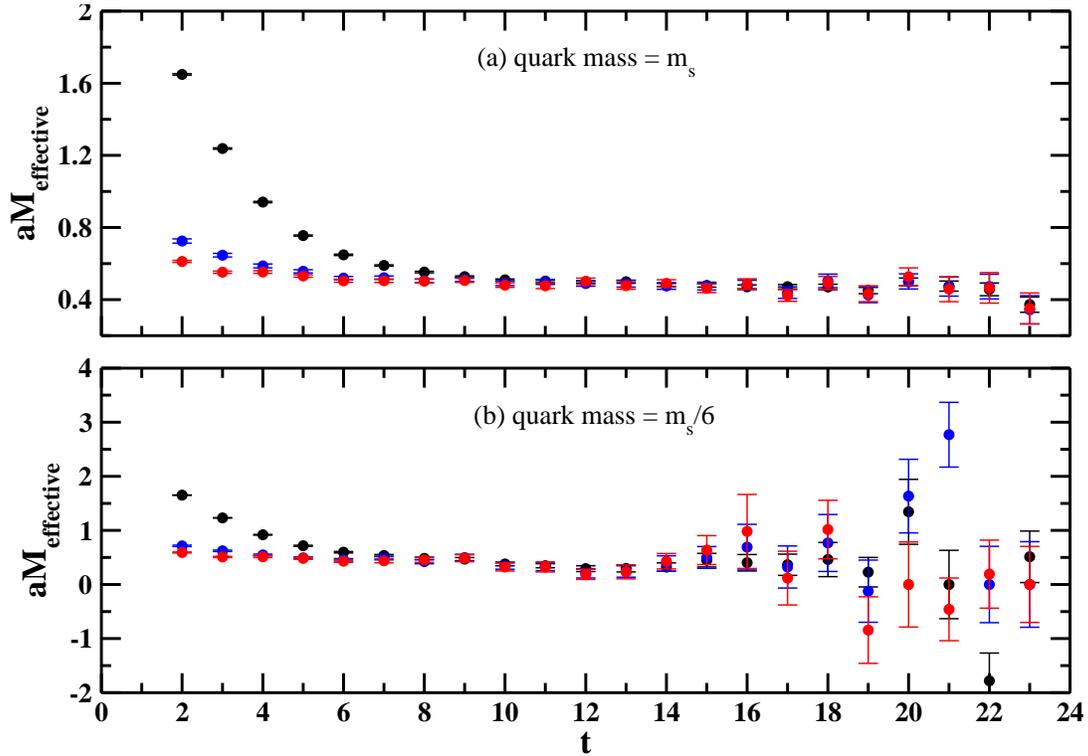}}
\end{center}
\caption{Effective mass plot for the charged vector at $\beta=6.0$ with
the (a) heaviest and (b) lightest quark masses.  Unsmeared results (black),
quark smeared results (blue), and quark and link smeared results (red) are
shown.}\label{quark-vs-link-smearing-effect}
\end{figure}

Gaussian quark smearing is used at the sink\cite{Alford:1995dm},
\begin{equation}
q^{\rm smr}(x)=\left(1+\alpha \Delta\right)^{n_\alpha}q(x),
\end{equation}
\begin{equation}
\Delta q(x) = \sum_{j=\pm1,\pm2,\pm3}\left[U_j(x)q(x+\hat{j})-q(x)\right].
\end{equation}
with stout spatial link variables\cite{Morningstar:2003gk}

%[Morningstar and Peardon, Phys. Rev. D69, 054501 (2004)],
\begin{equation}
U\rightarrow U^{(1)}\rightarrow U^{(2)} \rightarrow \dots \rightarrow U^{(n_\rho)},
\end{equation}
\begin{equation}
U_j^{(n+1)}(x)=\exp\left(i\rho \Theta_j^{(n)}(x)\right)U_j^{(n)}(x).
\end{equation}
The four smearing parameters were optimized
for the reduction of excited state contributions at $\beta$=6.0,
leading to $\alpha$=0.15, $n_\alpha$=64, $\rho$=0.15 and $n_\rho$=16. The link smearing
parameters were optimized first using the static quark anti-quark potential (see Figure
\ref{qqbar-potential} and compare with \cite{Basak:2005gi}). A comparison between the
effect of quark smearing and link smearing in reducing excited state contributions at our
lightest and our heaviest masses is shown in Figure \ref{quark-vs-link-smearing-effect}.

\section{Results}
\subsection{Vector Mesons}
In \cite{Abdel-Rehim:2006ve}, results for the pseudoscalar and vector mesons with local 
operators were reported.
A mass splitting between charged and neutral kaons was observed but, because of 
the large errors, no similar splitting between the neutral and charged
vectors was found.

\begin{figure}
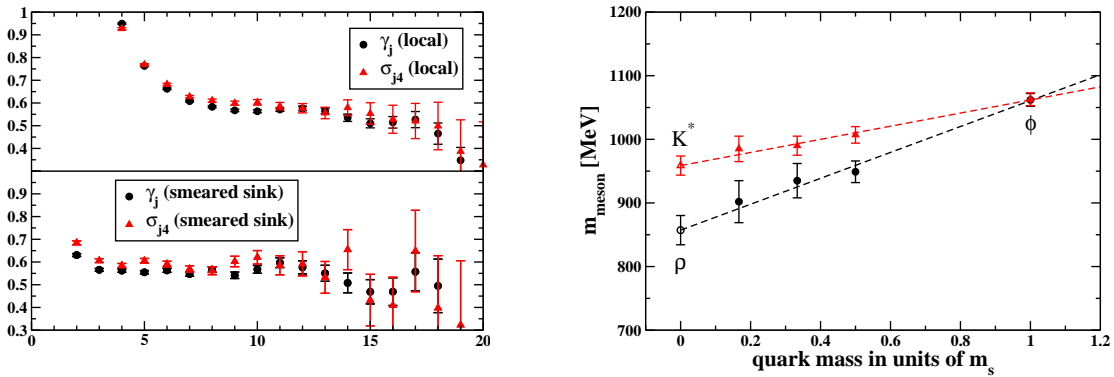
        
\begin{center}
\scalebox{0.28}{\includegraphics*[15mm,10mm][25cm,190mm]{localsmear.eps}}
\hspace{1cm}
\scalebox{0.28}{\includegraphics*[1mm,10mm][255mm,190mm]{mesonextrap.eps}}
\end{center}
\caption{The left panel shows effective mass plots for the neutral vector
meson with
$\beta=5.85$ and quark mass $=m_s/2$ for two local operators.
The right panel shows simple linear fits to vector meson masses after
continuum extrapolation.}
\label{vectors} 
\end{figure}

For any quark flavor $q$, the operators $\bar{q}\gamma_jq$ and
$\bar{q}\sigma_{j4}q$ both couple to the neutral vector meson,
but the latter can mix with the axial vector away from maximal twist.
In the present work, sink smearing was implemented for both of these
operators, and the charged-neutral splitting was still insignificant at
these lattice spacings.
Representative effective mass plots are shown in the left panel of
Figure \ref{vectors}.

Continuum extrapolations were performed linearly in $a^2$ for each
quark mass.  Subsequently, a simple linear fit in quark mass was used
to approximate a chiral extrapolation, as shown in the right panel of
Figure \ref{vectors}.  Masses for the $\rho$ and $K^*$ that were obtained
in this way are $O(10\%)$ above experiment, as expected for a quenched
simulation.

\subsection{Spin 1/2 baryons}
\begin{figure}
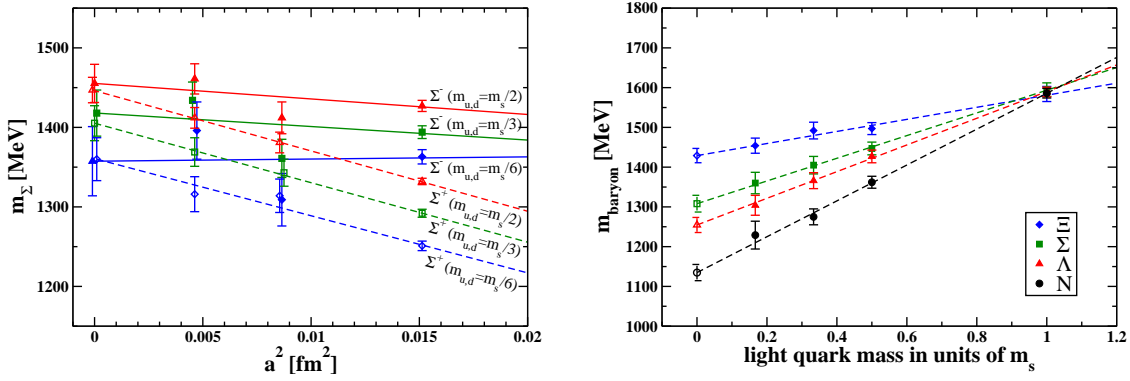

\begin{center}
\scalebox{0.28}{\includegraphics*[0mm,10mm][255mm,190mm]{sigmacompare.eps}}
\hspace{5mm}
\scalebox{0.28}{\includegraphics*[0mm,10mm][255mm,190mm]{spinhalfextrap.eps}}
\end{center}
\caption{The left panel shows scaling of the $\Sigma^+$ and $\Sigma^-$ masses
for three light quark masses, as well as linear extrapolations in $a^2$.
Tiny horizontal offsets are for clarity.
The right panel shows simple linear fits to the octet baryon masses, after
continuum extrapolations.}\label{spin-half-baryons}
\end{figure}

Source smearing allows for a more precise determination of ground state
octet baryon masses.  (For our previous exploration with local operators,
see \cite{Abdel-Rehim:2005qv}.) 
%{ Abdel-Rehim, Lewis and Woloshyn,PoS LAT2005:032,2006}.)
{}From the left panel of Figure \ref{spin-half-baryons}, one sees the
good scaling of the $\Sigma^-$ mass with lattice spacing,
while the $\Sigma^+$ has a significant linear dependence on $a^2$.
Since our tmLQCD corresponds to degenerate up and down quark masses, the
observed splitting among $\Sigma^\pm$ masses at $\beta=5.85$ is a twist
artifact.  For $\beta\geq6.0$, this artifact is not statistically
significant.  A similar twist artifact was observed among the $\Xi$ baryons.

Like the vector mesons, the quenched baryon masses obtained from simple
linear fits in the quark mass are systematically larger than
experiment, as evidenced by the right panel of Figure \ref{spin-half-baryons}.
The ordering of $N$, $\Lambda$, $\Sigma$ and $\Xi$ masses is correct.

\subsection{Spin 3/2 baryons}
\begin{figure}[t]
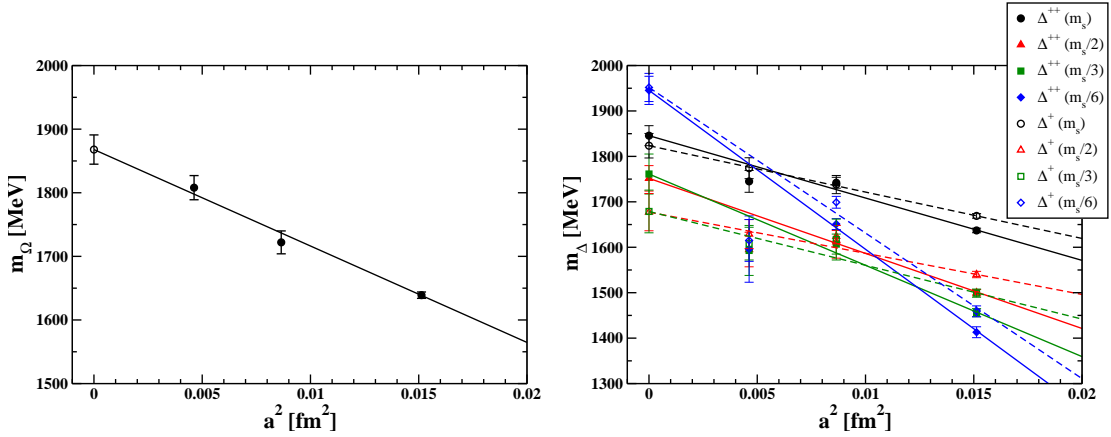

\begin{center}
\scalebox{0.28}{\includegraphics*[2mm,10mm][255mm,190mm]{omega.eps}}
\hspace{1mm}
\scalebox{0.28}{\includegraphics*[2mm,10mm][265mm,220mm]{delta.eps}}
\end{center}
\caption{Scaling of the $\Omega$ and $\Delta$ masses.  Linear fits to the
$\Delta$ data are illustrative; note that $\beta=6.0$ and 6.2 scale within
statistics.  Tiny horizontal offsets are for
clarity.}\label{spin-three-halfs-baryons}
\end{figure}

Even with smearing, quarks much lighter than $m_s$ make fitting
difficult. In Figure \ref{spin-three-halfs-baryons}, the best case ($\Omega$ has only
strange valence quarks) and the worst case ($\Delta$ has only light valence quarks)
are shown.
As expected, the quenched $\Omega$ is heavier than experiment.
The $\Delta$ data are noisier than one would like;
clearly sink smearing and/or additional operators should be explored in
this context.  Hints of twist artifacts can be seen at $\beta=5.85$, but
the over-all poor quality of $\Delta(1232)$ data underscores the
importance of future improvements like group-theoretical operators and
source smearing.

\section{Discussion}
Sink smearing was effective in reducing excited state
contributions.  However, this did not always lead to a more precise
prediction for the ground state mass, compared with {\em multi-state fits}
to unsmeared data. Nevertheless, some improvement in reducing systematic
errors arising due to multi-exponential fits may have been realized.

Reasonably precise results for vector meson and 
spin 1/2 baryon masses were obtained, and indications of twisting artifacts 
were identified on our coarsest lattice.  Even with smearing, the spin 3/2 
data were of low quality for $m_{u,d}<m_s/2$.
Negative parity baryons and excited states were included in the multi-state
fits, but were not useful for studies of the corresponding physical states and
are not shown here.

Sink smearing is useful but not sufficient to allow
precise studies of the baryon spectrum with tmLQCD.
In future studies, source smearing should be considered.  As well,
operators constructed according to the lattice symmetries are being
developed for tmLQCD\cite{tmgroup}.

\section*{Acknowledgements}
This work was supported in part by the Natural Sciences and Engineering Research Council
of Canada, the Canada Foundation for Innovation, the Canada Research Chairs Program and the
Government of Saskatchewan.


\begin{thebibliography}{99}

%\cite{Frezzotti:2000nk}
\bibitem{Frezzotti:2000nk}
  R.~Frezzotti, P.~A.~Grassi, S.~Sint and P.~Weisz  [Alpha collaboration],
  %``Lattice QCD with a chirally twisted mass term,''
  JHEP {\bf 0108}, 058 (2001)
  [arXiv:hep-lat/0101001].
  %%CITATION = HEP-LAT 0101001;%%

%\cite{Frezzotti:2003ni}
\bibitem{Frezzotti:2003ni}
  R.~Frezzotti and G.~C.~Rossi,
  %``Chirally improving Wilson fermions. I: O(a) improvement,''
  JHEP {\bf 0408}, 007 (2004)
  [arXiv:hep-lat/0306014].
  %%CITATION = HEP-LAT 0306014;%%

%\cite{Abdel-Rehim:2005gz}
\bibitem{Abdel-Rehim:2005gz}
  A.~M.~Abdel-Rehim, R.~Lewis and R.~M.~Woloshyn,
  %``Spectrum of quenched twisted mass lattice QCD at maximal twist,''
  Phys.\ Rev.\ D {\bf 71}, 094505 (2005)
  [arXiv:hep-lat/0503007].
  %%CITATION = HEP-LAT 0503007;%%

%\cite{Jansen:2005}
\bibitem{Jansen:2005}
  K.~Jansen, M~Papinutto, A.~Shindler, C.~Urbach and I.~Wetzorke,
  Phys.\ Lett.\ {\bf B619}, 184 (2005)
  [arXiv:hep-lat/0503031].
  %%CITATION = HEP-LAT 0503031;%%


%\cite{Abdel-Rehim:2005qv}
\bibitem{Abdel-Rehim:2005qv}
  A.~M.~Abdel-Rehim, R.~Lewis and R.~M.~Woloshyn,
  %``The hadron spectrum from twisted mass QCD with a strange quark,''
  PoS {\bf LAT2005}, 032 (2006)
  [arXiv:hep-lat/0509056].
  %%CITATION = HEP-LAT 0509056;%%

%\cite{Pena:2004gb}
\bibitem{Pena:2004gb}
  C.~Pena, S.~Sint and A.~Vladikas,
  %``Twisted mass QCD and lattice approaches to the Delta(I) = 1/2 rule,''
  JHEP {\bf 0409}, 069 (2004)
  [arXiv:hep-lat/0405028].
  %%CITATION = HEP-LAT 0405028;%%

%\cite{Farchioni:2004ma}
\bibitem{Farchioni:2004ma}
  F.~Farchioni {\it et al.},
  %``Exploring the phase structure of lattice QCD with twisted mass quarks,''
  Nucl.\ Phys.\ Proc.\ Suppl.\  {\bf 140}, 240 (2005)
  [arXiv:hep-lat/0409098].
  %%CITATION = HEP-LAT 0409098;%%

%\cite{Farchioni:2004fs}
\bibitem{Farchioni:2004fs}
  F.~Farchioni {\it et al.},
  %``The phase structure of lattice QCD with Wilson quarks and  renormalization
  %group improved gluons,''
  Eur.\ Phys.\ J.\ C {\bf 42}, 73 (2005)
  [arXiv:hep-lat/0410031].
  %%CITATION = HEP-LAT 0410031;%%

%\cite{Alford:1995dm}
\bibitem{Alford:1995dm}
  M.~G.~Alford, T.~Klassen and P.~Lepage,
  %``The D234 action for light quarks,''
  Nucl.\ Phys.\ Proc.\ Suppl.\  {\bf 47}, 370 (1996)
  [arXiv:hep-lat/9509087].
  %%CITATION = HEP-LAT 9509087;%%

%\cite{Morningstar:2003gk}
\bibitem{Morningstar:2003gk}
  C.~Morningstar and M.~J.~Peardon,
  %``Analytic smearing of SU(3) link variables in lattice QCD,''
  Phys.\ Rev.\ D {\bf 69}, 054501 (2004)
  [arXiv:hep-lat/0311018].
  %%CITATION = HEP-LAT 0311018;%%

%\cite{Basak:2005gi}
\bibitem{Basak:2005gi}
  S.~Basak {\it et al.},
  %``Combining quark and link smearing to improve extended baryon operators,''
  PoS {\bf LAT2005}, 076 (2006)
  [arXiv:hep-lat/0509179].
  %%CITATION = HEP-LAT 0509179;%%

%\cite{Abdel-Rehim:2006ve}
\bibitem{Abdel-Rehim:2006ve}
  A.~M.~Abdel-Rehim, R.~Lewis, R.~M.~Woloshyn and J.~M.~S.~Wu,
  %``Strange quarks in quenched twisted mass lattice QCD,''
  Phys.\ Rev.\ D {\bf 74}, 014507 (2006)
  [arXiv:hep-lat/0601036].
  %%CITATION = HEP-LAT 0601036;%%

\bibitem{tmgroup} D.~Harnett, R.~Lewis and R.~G.~Petry, these proceedings
  [arXiv:hep-lat/0609071].

\end{thebibliography}
\end{document}